# 新冠状病毒感染在湖北省扩散的 SIR 模型预测

SIR Model-based Prediction of Infected Population of Coronavirus in Hubei Province

张钧 [1]　王立洪 [2]　王骥 [1,*]

[1] 宁波大学机械工程与力学学院力学与工程科学系

[2] 宁波大学数学与统计学院数学系

[*] 通讯作者：wangji@nbu.edu.cn

## 摘 要

武汉新冠状病毒（Coronavirus）疫情突然爆发后，连续完整的公开感染数据成为公众了解疫情的重要信息，也是政府防控决策的重要依据，对于及时制定防控策略和落实措施有重要意义。随着疫情的发展和扩散，未来的感染数据对于后续的防控策略和资源配置更加重要，需要深入细致的研究，提供尽可能精确可靠的预测数据。这里选择常用的 SIR 感染模型并结合公布的湖北省实际疫情数据，通过非线性拟合和误差平方和最小原则确定关键参数，对疫情的发展过程提供了一个精确可靠的系统估计。

关键词：湖北；新冠状病毒；感染；病人；预测

## Abstract

After the sudden outbreak of Coronavirus in Wuhan, continuous and rich data of the epidemic has been made public as the vital fact for decision support in control measures and aggressive implementation of containment strategies and plans. With the further growth and spreading of the virus, future resource allocation and planning under updated strategies and measures rely on careful study of the epidemic data and characteristics for accurate prediction and estimation. By using the SIR model and reported data, key parameters are obtained from least sum of squared errors for an accurate prediction of epidemic trend in the last four weeks.

## 1. 武汉冠状病毒疫情

武汉新冠状病毒（Coronavirus, 2019-nCov, Covid-19, NCP, SARS-Cov-2）爆发的公开宣布在 2019 年 1 月 21 日，随后大量的疫情数据开始公布[1, 2]，显示出了疫情扩散的严重性，引起了世界性的高度关注。此后一连串的疫情控制措施，如加强医疗救治、隔离、强化医疗资源、研究诊治方法和疫苗等，以及后来的封城等防控策略，都和未来的疫情发展有关。目前疫情还在

发展，全国性的救援和控制行动正在更大的范围展开，未来疫情发展的趋势预测也在更大的范围对疫情的控制发挥着关键作用[3-7]。

传染病的疫情扩散模型是一个传统的流行病学和数学问题，有着重要的实用价值。近年来世界范围内的大规模疫情频繁爆发，更引起学术界和社会各界对传染模型及其应用的高度关注，在方法上也有了一些创新，更在疫情控制、救助行动、资源调配、交通运输、旅行规划等方面发挥着重要作用。在全世界开展的智能城市建设等技术创新活动中，结合城市核心功能开展传染病扩散模型研究，可以深化数字城市模型、提高管理水平、完善应急能力，是城市建设和发展的重要内容。这方面的研究能在不断成熟的基础上，和未来的城市和社会发展模型结合，提升城市应急管理水平，保障人类可持续发展目标的最终实现。

目前有众多的传染病扩散模型和分析方法。本文从简单的 SIR 模型出发，基于湖北省近期公布的感染人数，利用误差平方和(SSE)最小方法确定了两个模型参数并进行了验证，确认模型预测数据和实际公布的疫情数据非常接近[8]。以此模型为基础，我们对未来的疫情发展进行了预测，结果和后续其他预测一致，提前确认了 2 月 19 日为现存感染人数的拐点。由于疫情总体趋势预测的重要性，我们采用的简单方法的预测结果和实际公布数据一致，准确预测了疫情发展态势，对于疫情研究和防控工作都有重要意义。我们也和其他的感染传播数学模型的结果进行了比较，确认在疫情预测方面这一方法非常直观准确。

## 2. SIR 感染模型与方法介绍

传染病是一个长期存在的社会和自然现象，在人类历史上成为了一些重要改变的转折点或者巨大变化的触发点。也就是由于这些原因，传染病的扩散模型也是我们熟悉的，也是人类在传染病防控方面的重要成就之一，经常在新闻和专题

报道中提到。作为一个社会现象，研究上通常采取不少的假设，但它们的有效性和合理性已经得到证实，在今天全球传染病防控方面发挥着重要作用。这里我们采用尽可能简单的模型，是为了满足当前对感染人数和疫情总体趋势的迅速估计，帮助制定精准的病人救助措施和疫情控制策略研究。考虑人工干预的模型会更加复杂，需要收集和分析更多的数据，但可以在我们研究的基础上继续开展工作。

我们熟悉的 SIR 模型是基于疫情流行区域的总人数$N$、感染人数$I$、易感人数$S$、病愈人数$R$和时间 $t$之间的如下关系：

$$\frac{dS}{dt} = -\beta \frac{IS}{N},$$
$$\frac{dI}{dt} = \beta \frac{IS}{N} - \gamma I, \qquad (1)$$
$$\frac{dR}{dt} = \gamma I,$$

这些方程里的参数$\beta$和$\gamma$为常数，反映了特定疫情的特征。这些方程貌似简单，但由于常数$\beta$和$\gamma$是同一数量级，导致方程属于高度耦合的非线性类型，实际上无法求解析解，需要用数值解来提供预测结果。

由于方程求解上的困难，我们通常采取数值解法等方法来求解。事实上，数值解法是这类问题研究工作常用的方法，效果也很好。至于其他的方法如解析法[9]，目前有些研究工作，但由于过程复杂，应用会受到限制。这里我们在数值解的基础上开展参数估计方面的研究工作。

在疫情扩散过程中的早期，由于开始时易感人群也就是总人数，即$S \approx N$，我们可以简化感染人数$I$和时间$t$的关系为

$$\frac{dI}{dt} = \beta \frac{S}{N} I - \gamma I \approx (\beta - \gamma)I. \qquad (2)$$

由此可得到感染人数的近似解为

$$I(t) = e^{(\beta-\gamma)t}. \tag{3}$$

这一关系表明，近似的感染人数总数是时间的指数函数。这里的常数$\beta$和$\gamma$应该根据疫情的特点来确定，从而实现感染人数的估计。当然，疫情防控措施也会影响这些参数，反过来也反映了防控措施的效果。这些参数一般是根据流行病学的统计结果得到的，会在疫情的流行过程中得到反映。也就是说，我们也可以根据实际疫情报告来决定这些参数。由于我们已经积累了一些疫情实际数据，基于SIR分析的回溯拟合可以精确地确定这些参数。

但是，我们发现，基于方程（3）的简单模型可以大致地确定这些参数，可以在疫情初期评估疫情的特征，但简单的指数增长模型不能反映疫情防控的全部过程。很明显，更可靠的估计应该从实际疫情估计参数，用 SIR 完整模型进行计算。

首先，根据早期的公开数据对方程（3）进行最小二乘法模拟，我们得到初始的参数估计值为$\beta$=3.7045 和 $\gamma$=3.5250。利用这些参数和方程（3），我们发现早期的实际感染人数和估计值一致性非常好。也就是说，在疫情的暴发期，近似的 SIR 模型解可以简单且准确地估计感染人数。

获得了这两个参数的初始值之后，我们的目标就是在初始值的基础上，利用方程（1）的数值解和实际感染数据，找到基于 SIR 模型估计的最佳参数。微分方程（1）可以用 Runge-Kutta 法求解，其结果当然依赖于参数$\beta$和$\gamma$。我们利用初始值设定一个搜索区间，然后找到计算的感染人数和公布的感染人数的误差之和最小的数值，就可以据此来计算疫情的趋势。也就是说，我们的思路是在SIR模型中选取使预测值最接近公开数据的$\beta$和$\gamma$参数，利用SIR模型预测疫情。具体地来说，我们的计算流程是[8]：

1) 给定疫情初始参数：β=3.7045，γ=3.5250；
2) 用 Runge-Kutta 法从方程（1）计算与实际感染人数$I_r$对应的预测感染人数$I_c$；
3) 用$M$个实际数据计算误差平方和（SSE）$\Delta = \sum_{i=1}^{M}\left(I_c^{(i)} - I_r^{(i)}\right)^2$；
4) 选择使Δ值最小的参数β和γ为疫情特征参数。

我们可以用这里获得的疫情参数和方程（1）预测疫情的发展趋势。当然，数值的计算仍然需要使用 Runge-Kutta 法。

## 3. 基于实际确诊数据的模型参数估计

这次武汉新冠状病毒肺炎爆发后，疫情数据从 2020 年 1 月 24 日开始每天定期公布，有完整的数据可以根据我们前面提出的数学模型来完成参数估计。这些计算相对简单，但是涉及到一个疫情数据选择的问题。由于疫情从武汉市开始，初期的各种环境和湖北省完全一致，我们决定选择湖北全省的疫情数据对前述的 SIR 模型进行模拟。这些计算相对简单，有许多现成的软件可以使用，具体过程我们就不做详细介绍了。我们计算的参数初值为β=3.7045 和 γ= 3.5250，在计算过程中发现这些选择是合理的。

我们用简单的 SIR 模型对武汉冠状病毒的初期感染确诊数据进行了回溯分析，按不同实际时间长度的拟合结果见图 1-4 和表 1，可以发现实际数据和 SIR 预测数据高度一致，有助于我们准确认识冠状病毒的特征和早期传播规律等，也对于指导以后的干预策略有重要意义。注意这里的确诊数据是指现存的确诊人数，它会有一个上升期和下降期，中间有明显的峰值，也就是常说的转折点。利用这些较为充分的数据，可以更准确地提供一些关键参数如潜伏期和感染系数等。这部分的工作也是更完整的 SEIR 模型分析的基础。

从这里的结果可以看出，实际数据符合 SIR 模型，这一点是长期一致的。不同时长拟合的参数有区别，反映了疫情特征的变化和数据的复合效应。几个显著的特点是：

1) 感染人数和拐点随着实际数据样本而增加。也就是说，早期的预测不能反映疫情全貌。

2) 这一方法能够很好地提前预测今后三天的感染人数。

3) 利用这一方法预测的累计感染人数（$I+R$）在峰值之前也很可靠。

4) 当实际数据接近疫情拐点时，从第二十天（2月14日），预测峰值结果稳定地收敛于 2 月 17 日，而实际数据是 18 日。

5) 这里预测的拐点和实际峰值一致，再次证明 SIR 模型是可靠的。

6) 疫情总体趋势的预测不必采取分段拟合的方式。整体数据能够很好地反映疫情的关键特征。

7) 疫情经过拐点后，后续的数据不会改变疫情的拐点位置。

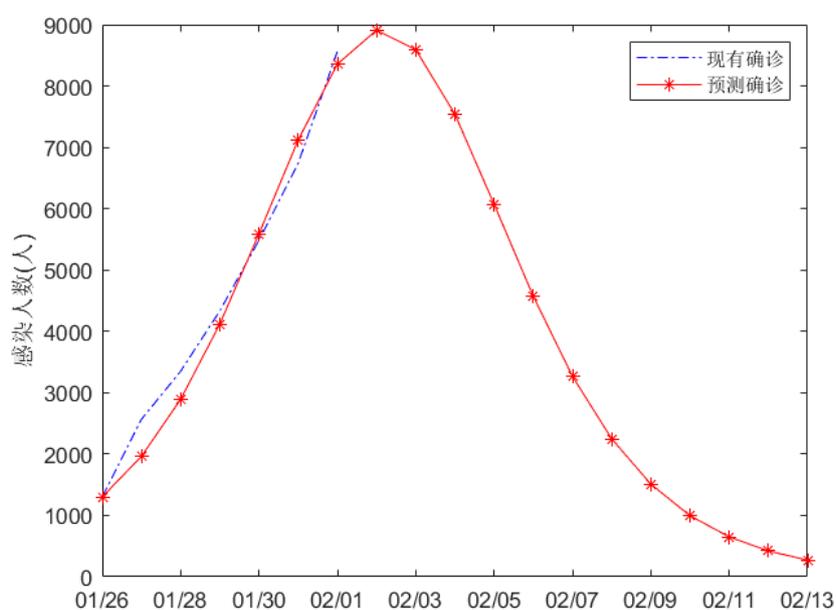

图1 湖北省前7天的疫情扩散数值模拟结果(β=25.4396和γ= 25.0215)

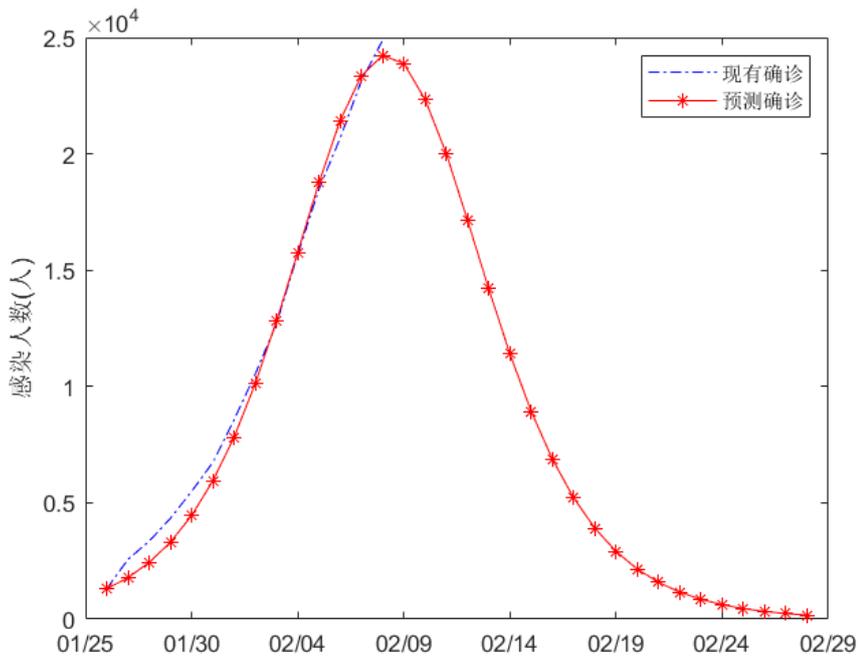

图 2 湖北省前 14 天的疫情扩散数值模拟结果(β= 11.0842 和γ= 10.7685)

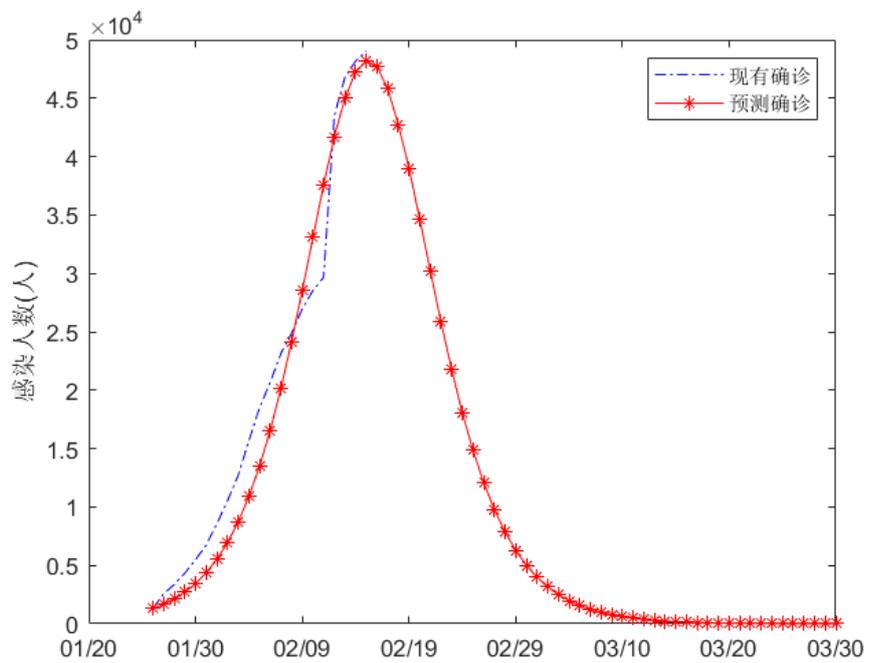

图 3 湖北省前 21 天的疫情扩散数值模拟结果(β= 6.0163 和γ=5.7719)

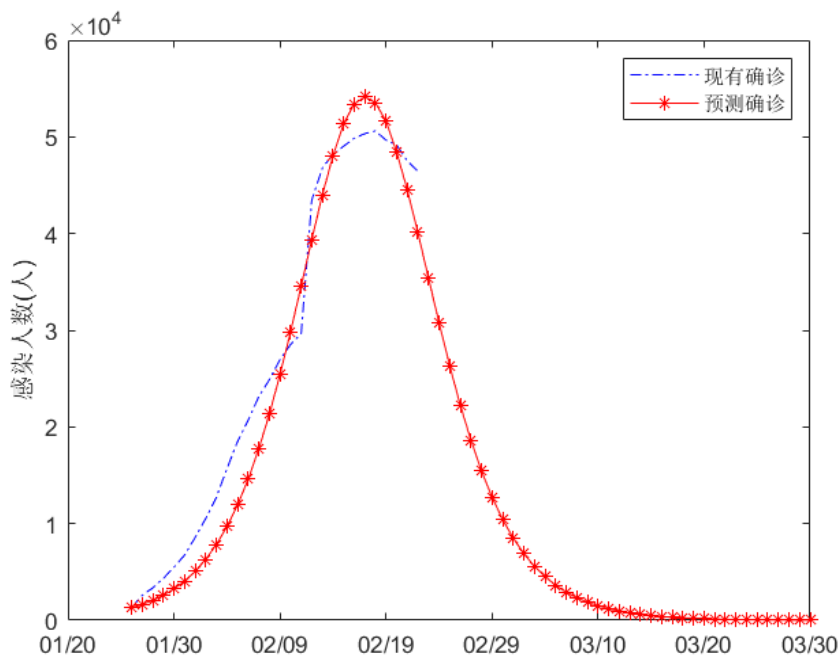

图 4 湖北省前 28 天的疫情扩散数值模拟结果(β = 5.3130 和 γ = 5.0839)

表 1 湖北省前 28 天的数值模拟结果与实际数据对照表(β= 5.3130 和 γ= 5.0839)

| 日期 | 预测确诊 | 现有确诊 | 日期 | 预测确诊 | 现有确诊 | 日期 | 预测确诊 | 现有确诊 | 日期 | 预测确诊 | 现有确诊 |
|---|---|---|---|---|---|---|---|---|---|---|---|
| 1.26 | 1303 | 1303 | 2.14 | 48042 | 48175 | 3.05 | 5574 | | 3.24 | 81 | |
| 1.27 | 1638 | 2567 | 2.15 | 51355 | 49030 | 3.06 | 4496 | | 3.25 | 65 | |
| 1.28 | 2058 | 3349 | 2.16 | 53410 | 49847 | 3.07 | 3619 | | 3.26 | 52 | |
| 1.29 | 2584 | 4334 | 2.17 | 54137 | 50338 | 3.08 | 2909 | | 3.27 | 41 | |
| 1.30 | 3239 | 5486 | 2.18 | 53529 | 50620 | 3.09 | 2335 | | 3.28 | 33 | |
| 1.31 | 4053 | 6738 | 2.19 | 51608 | 49664 | 3.10 | 1872 | | 3.29 | 26 | |
| 2.01 | 5062 | 8565 | 2.20 | 48427 | 49156 | 3.11 | 1500 | | 3.30 | 21 | |
| 2.02 | 6306 | 10532 | 2.21 | 44560 | 47535 | 3.12 | 1200 | | 3.31 | 17 | |
| 2.03 | 7831 | 12712 | 2.22 | 40131 | 46418 | 3.13 | 960 | | | | |
| 2.04 | 9685 | 15679 | 2.23 | 35423 | | 3.14 | 768 | | | | |
| 2.05 | 11921 | 18483 | 2.24 | 30791 | | 3.15 | 614 | | | | |
| 2.06 | 14585 | 20677 | 2.25 | 26344 | | 3.16 | 490 | | | | |
| 2.07 | 17712 | 23139 | 2.26 | 22285 | | 3.17 | 392 | | | | |
| 2.08 | 21319 | 24881 | 2.27 | 18643 | | 3.18 | 313 | | | | |
| 2.09 | 25389 | 26965 | 2.29 | 15466 | | 3.19 | 250 | | | | |
| 2.10 | 29857 | 28532 | 3.01 | 12735 | | 3.20 | 200 | | | | |
| 2.11 | 34579 | 29659 | 3.02 | 10424 | | 3.21 | 159 | | | | |
| 2.12 | 39375 | 43455 | 3.03 | 8493 | | 3.22 | 127 | | | | |
| 2.13 | 43998 | 46806 | 3.04 | 6892 | | 3.23 | 102 | | | | |

# 4. 结论

通过利用 SIR 模型对实际疫情数据的拟合，我们成功获得了疫情主要参数，完整准确地预测了疫情的发展趋势。后来的实际疫情数据表明，我们利用这些数据成功预测了感染病人的高峰，也就是常提到的所谓拐点，会在 2 月 17 日出现，这和后来的实际数据的 2 月 18 日基本完全吻合。这些结果表明，本文的简单计算方法和累积的实际数据可以在拐点之前准确地预测疫情走势，为疫情防控提供可靠的科学依据。当然，拐点之后的预测也比较好，对于研究疫情后期的控制方案有重要参考意义。基于简单的 SIR 模型对这次武汉新花冠病毒的传播规律进行了回溯分析结果表明，再次确认 SIR 模型在传染病扩散模式预测上是可靠的。这一模型和方法能够获得较为可靠地确定病毒感染特征数据，为制定今后的疫情干预决策提供数据支持，保证社会和医疗资源配置的最大效益。